\documentclass[final,5p,times,twocolumn]{elsarticle}
\usepackage{url}
\usepackage{graphicx}
\usepackage{amssymb}
\usepackage{lineno}
\biboptions{square,comma,numbers,sort&compress}
\journal{Nucl. Instrm. Methods Phys. Res. Section A}
\begin{document}
\begin{frontmatter}
\title{DALI2: A NaI(Tl) detector array for measurements of $\gamma$ rays from fast nuclei}
\author[riken]{S.~Takeuchi\corref{cor1}}
\cortext[cor1]{Corresponding author.}
\ead{takesato@riken.jp}
\author[riken]{T.~Motobayashi}
\author[titech]{Y.~Togano}
\author[cns]{M.~Matsushita}
\author[rcnp]{N.~Aoi}
\author[rikkyo]{K.~Demichi}
\author[rikkyo]{H.~Hasegawa}
\author[riken]{H.~Murakami}

\address[riken]{RIKEN Nishina Center, Wako, Saitama 351-0198, Japan}
\address[titech]{Department of Physics, Tokyo Institute of Technology, Megro, Tokyo 152-8551, Japan}
\address[cns]{CNS, University of Tokyo, RIKEN campus, Wako, Saitama 351-0198, Japan}
\address[rcnp]{RCNP, Osaka University, Mihogaoka, Ibaraki, Osaka, 567-0047, Japan}
\address[rikkyo]{Department of Physics, Rikkyo University, Toshima, Tokyo 171-8501, Japan}

\begin{abstract}
A NaI(Tl) detector array called DALI2 (Detector Array for Low Intensity radiation 2) 
has been constructed for in-beam $\gamma$-ray spectroscopy experiments with fast radioactive isotope (RI) beams.
It consists typically of 186 NaI(Tl) scintillators covering polar angles 
from $\sim$15$^{\circ}$ to $\sim$160$^{\circ}$ with an average angular resolution of 6$^{\circ}$ 
in full width at half maximum. 
Its high granularity (good angular resolution) enables Doppler-shift corrections that result in, 
for example, 10\% energy resolution and 20\% full-energy photopeak efficiency for 1-MeV $\gamma$ rays 
emitted from fast-moving nuclei (velocities of $v/c \simeq 0.6$). 
DALI2 has been employed successfully in numerous experiments using fast RI beams 
with velocities of $v/c = 0.3 - 0.6$ provided by the RIKEN RI Beam Factory. 
\end{abstract}


\end{frontmatter}


\section{Introduction}
\label{intro}

The introduction of large arrays to $\gamma$ detection systems 
for nuclear spectroscopy was primarily aimed at coping with heavy-ion induced reactions, 
which may produce cascades of $\gamma$ rays containing multiple transitions.
The Darmstadt-Heidelberg Crystal Ball, composed of 76 NaI(Tl) crystals,  
was a pioneering array developed in the 1980's~\cite{metag}.
Its ability to measure high-multiplicity events with high detection efficiency 
enabled the identification of states 
with higher spins compared to earlier studies 
that used conventional setups. 

One direction of development was toward higher energy-resolution arrays 
by utilizing germanium (Ge) detectors~\cite{gammasphere,euroball,exogam,sega,greta1,greta2,grape,agata}.
GAMMASPHERE~\cite{gammasphere} and EUROBALL~\cite{euroball} are 
examples of large-scale arrays used in various spectroscopic experiments. 
Recent developments along this line are Ge tracking arrays such as GRETA~\cite{greta1,greta2} and AGATA~\cite{agata}.  
Another direction is toward higher efficiency measurements. 
The BaF$_2$-crystal array was developed for astrophysical (n,$\gamma$) studies at Karlsruhe~\cite{karlsruhe}.  
Another example CAESAR based on CsI(Na) scintillators is optimized for high $\gamma$-ray detection efficiency 
in in-beam $\gamma$-ray spectroscopy experiments 
at the National Superconducting Cyclotron Laboratory 
at Michigan State University~\cite{caesar}. 
For low-energy radioactive-beam experiments, the Miniball spectrometer using Ge detectors has been operational at REX-ISOLDE at CERN~\cite{miniball}.

The recent development of fast radioactive ion beams
requires a new type of $\gamma$-ray detector array for in-beam spectroscopic studies.
Since experiments with such beams are performed in inverse kinematics 
where the $\gamma$-ray emitter has high velocity, typically in the range $v/c \simeq$ 0.3 to 0.6,  
causing a large Doppler shift, 
$\gamma$-ray measurements require information on the direction of the radiation 
to extract the transition energy in the rest frame of the projectile. 
High efficiency is another requirement, since the secondary-beam intensity for 
nuclei far from stability is typically low.
The CAESAR mentioned above is one of the arrays for such fast-beam applications.
The RIKEN Radioactive Isotope Beam Factory (RIBF)~\cite{RIBF} 
provides unique opportunities to research unstable nuclei 
owing to the production of the world's highest intensity exotic beams available today.
To fully exploit the performance of the facility, 
we have built a new $\gamma$-ray detector array called DALI2.
It consists of 186 NaI(Tl) crystals in its standard configuration, 
and has a reasonably good angular resolution and a high detection efficiency.

The design of DALI2 follows a similar concept to the original array 
DALI (Detector Array for Low Intensity radiation)~\cite{NSR1995MO16,dali},  
which was developed for experiments at the old facility at RIKEN 
that provides light exotic beams with $v/c \approx 0.3$.
DALI has been successfully employed, 
owing to its versatility with respect to detector configuration and 
its reasonably good angular and energy resolutions, 
in various in-beam $\gamma$-ray spectroscopy measurements~\cite{NSR1995MO16, NSR1997NA08, NSR2000IW02, NSR2000IW03, NSR2001YO03, NSR2001IW07, NSR2003IW02, NSR2003SH06, NSR2003YA05, NSR2004YA01, NSR2004EL03, NSR2004IM01, NSR2005KO13, NSR2005IW02, NSR2008IW04, NSR2009KO02} (see Table~\ref{explist1}). 
However, for higher-velocity beams, 
the performance of DALI is not optimized. 
DALI2 was designed to 
meet the criteria for the new RIBF facility, 
where the RI-beam velocity is typically 60\% of the light speed, 
by improving the angular resolution 
and the detection efficiency. 
DALI2 was completed in 2002, and was used in a variety of experiments 
at the old RIKEN facility and more recently at the new facility. 
Gamma-transitions associated with direct reactions  
including Coulomb excitation, proton and alpha inelastic scattering, 
secondary fragmentation, and nucleon transfer reactions~\cite{NSR2004EL10, NSR2005KA06, NSR2005EL07, NSR2005DO16,NSR2006ON02, NSR2006MI16, NSR2006EL05, NSR2006EL03, NSR2006DO09,NSR2007GI06, NSR2008SU12, NSR2008ON02, NSR2008IW03, NSR2009EL03, NSR2009AO01, NSR2009TA08, NSR2009DO10, NSR2010DO05, NSR2010EL05, NSR2012TO06, NSR2012LI45, NSR2012TA20, NSR2013SU20, NSR2013ST20, NSR2013DO22} have been studied (see Table~\ref{explist2}).
This article focuses on the design and performance of DALI2.  
Features of DALI2 in its early stage are briefly reported in ref.~\cite{dali2,dali2-web}.

\section{Description of DALI2}
\subsection{Basic design}

The center-of-mass energy resolution for $\gamma$ rays emitted from moving sources depends on the uncertainty 
in the measurements of the angle of emission together with 
the velocity spread of the projectile and 
the intrinsic energy-resolution of the detector, 
which will be discussed in detail in Sect.~\ref{performance}. 
We adopted NaI(Tl) scintillator as the material for the detector crystal, 
since it offers a good compromise between intrinsic resolution, detection efficiency and costs.
The angle measurement is made by employing a large number of ($160-186$ depending on the experimental condition) 
detectors at various distances from the target.
The angle of $\gamma$-ray emission is also useful to characterize the $\gamma$ transition, 
because the $\gamma$-ray angular distribution is sensitive to the transition multipolarity ($\Delta L$),  
reflecting the nuclear spin-alignment produced during the reaction process 
that creates the nuclei being studied. 

The array should cover a wide range of angles, 
to be sensitive to $\Delta L$ and, more importantly, 
to reduce the uncertainty in extracting the total $\gamma$ emission yield. 
Especially, small-angle measurements are important, 
because the $\gamma$-ray angular distribution is forward peaked in the laboratory frame due to the Lorentz boost.
DALI2 can measure $\gamma$ rays from 15$^{\circ}$ 
to enable to cover forward-peaked $\gamma$-ray angular distribution 
for $\gamma$ emitters with the velocities around $v/c=0.6$ at the new RIBF facility.

\begin{table*}[hp]
\caption{List of experiments performed with DALI.}
\label{explist1}
\center{
\begin{tabular}{llccclc}
\hline
\hline
Reaction & Beam & Energy & Target & Thickness & Observables & Reference \\
         &      & (MeV/u)&        & (mg/cm$^{2}$) &         &           \\
\hline
Pb($^{32}$Mg, $^{32}$Mg+$\gamma$) & $^{32}$Mg & 49.2  & Pb & 350  & $B$(E2)~\footnotemark & ~\cite{NSR1995MO16} \\
Pb($^{11}$Be, $^{11}$Be+$\gamma$) & $^{11}$Be & 63.9  & Pb & 350  & d$\sigma$/d$\Omega$~\footnotemark, $B$(E1) & ~\cite{NSR1997NA08} \\
(CH$_{2}$)$_{n}$($^{10,12}$Be, $^{10,12}$Be+$\gamma$) & $^{10}$Be & 59.2 & (CH$_{2}$)$_{n}$ & 90.2 & d$\sigma$/d$\Omega$, $\delta$~\footnotemark & ~\cite{NSR2000IW02} \\
$^{12}$C($^{10,12}$Be, $^{10,12}$Be+$\gamma$) & $^{12}$Be & 53.8 & C & 89.8 & & \\
Pb($^{12}$Be, $^{12}$Be+$\gamma$) & $^{12}$Be & 53.3 & Pb & 350.8 & $E(1_{1}^{-})$~\footnotemark, $B$(E1) & ~\cite{NSR2000IW03} \\
$^{12}$C($^{12}$Be, $^{12}$Be+$\gamma$) &     & 54.0 & C & 89.8 & & \\
$^{9}$Be($^{36}$Si, $^{34}$Mg+$\gamma$) & $^{36}$Si & 38.0  & Be & 385  & $E(2_{1}^{+})$, $E(4_{1}^{+})$, $R(4^{+}/2^{+})$~\footnotemark & ~\cite{NSR2001YO03} \\
Pb($^{34}$Mg, $^{34}$Mg+$\gamma$) & $^{34}$Mg & 44.9  & Pb & 693  & $B$(E2) & ~\cite{NSR2001IW07} \\
$^{2}$H($^{34}$Si, $^{34}$Si+$\gamma$) & $^{34}$Si & 38.4  & Liq.D$_{2}$~\footnotemark & 150  & levels & ~\cite{NSR2003IW02} \\
$^{12{\rm m}}$Be$\rightarrow$$^{12}$Be$^{*}$(2$^{+}$)$\rightarrow$$^{12}$Be$_{\rm g.s.}$ & $^{12({\rm m})}$Be & 60.0  & - & - & $E(0_{2}^{+})$, $\gamma$-$\gamma$ angular correlation & ~\cite{NSR2003SH06} \\
$^{1}$H($^{30}$Ne, $^{30}$Ne+$\gamma$) & $^{30}$Ne & 48.0  & Liq.H$_{2}$ & 186  & $E(2_{1}^{+})$, $B$(E2) & ~\cite{NSR2003YA05} \\
Pb($^{15}$O, $^{15}$O+$\gamma$) & $^{15}$O & 85.0  & Pb & 1480  & $\Gamma_{\gamma}$ of the 3/2$^{+}$ state at 6.793 MeV & ~\cite{NSR2004YA01} \\
Pb($^{16}$C, $^{16}$C+$\gamma$) & $^{16}$C & 52.7  & Pb & 50  & d$\sigma$/d$\Omega$, $B$(E2) & ~\cite{NSR2004EL03} \\
$^{9}$Be($^{16}$C, $^{16}$C+$\gamma$) & $^{16}$C & 34.6  & Be & 370  & lifetime, $B$(E2) & ~\cite{NSR2004IM01} \\
$^{12}$C($^{15,17}$B, $^{15,17}$B+$\gamma$) & $^{15,17}$B & 72.0  & C & 377  & levels, d$\sigma$/d$\Omega$, $\delta$ & ~\cite{NSR2005KO13} \\
Pb($^{28}$Ne, $^{28}$Ne+$\gamma$)       & $^{28}$Ne & 46.0 & Pb & 693 & $B$(E2) & ~\cite{NSR2005IW02} \\
$^{12}$C($^{28}$Ne, $^{28}$Ne+$\gamma$) &           &      & C & 339 & & \\
Pb($^{20}$Mg, $^{20}$Mg+$\gamma$)       & $^{20}$Mg & 58.0 & Pb & 226 & d$\sigma$/d$\Omega$, $B$(E2) & ~\cite{NSR2008IW04} \\
$^{12}$C($^{20}$Mg, $^{20}$Mg+$\gamma$) &           &      & C & 118 & & \\
$^{1}$H($^{18}$C,$^{17}$C+$\gamma$) & $^{18}$C & 81.0 & Liq.H$_{2}$ & 120 & levels, $\sigma$~\footnotemark, d$\sigma$/d$P_{x}$~\footnotemark & ~\cite{NSR2009KO02} \\
$^{1}$H($^{19}$C,$^{18}$C+$\gamma$) & $^{19}$C & 68.0 &             &     & & \\
\hline
\hline
\end{tabular}
}
\begin{flushleft}
\hspace{1em}
\small{
\begin{tabular}{rl}
\\
1& $B$(EL/ML): transition probability for EL or ML transition\\
2& d$\sigma$/d$\Omega$: differential cross section\\
3& $\delta$: deformation length\\
4& $E(J^{\pi})$: energy of excited state\\
5& $R(4^{+}/2^{+})$: energy ratio between the $4^{+}$ and $2^{+}$ states\\
6& Liq.D$_{2}$, Liq.H$_{2}$: liquid targets of deuterium and hydrogen provided by CRYogenic ProTon and Alpha target system (CRYPTA)~\cite{crypta}\\
7& $\sigma$: total or angle integrated cross section\\
8& d$\sigma$/d$P_{x}$: parallel momentum distribution\\
\end{tabular}
}
\end{flushleft}
\end{table*}

\begin{table*}[hp]
\caption{List of experiments performed with DALI2.}
\label{explist2}
\center{
\begin{tabular}{llccclc}
\hline
\hline
Reaction & Beam & Energy & Target & Thickness & Observables & Reference \\
         &      & (MeV/u)&        & (mg/cm$^{2}$) &         &           \\
\hline
$^{1}$H($^{27}$F, $^{25,26,27}$F+$\gamma$) & $^{27}$F & 39.6  & Liq.H$_{2}$ & 210  & levels, $\sigma$ & ~\cite{NSR2004EL10} \\
$^{1}$H($^{17}$B, $^{12,14,15,17}$B+$\gamma$) & $^{17}$B & 43.0  & Liq.H$_{2}$ & 180  & levels, $\sigma$ & ~\cite{NSR2005KA06} \\
$^{1}$H($^{19}$C, $^{17,19}$C+$\gamma$) & $^{19}$C & 49.4 & Liq.H$_{2}$ & 190  & levels, $\sigma$, $\beta$~\footnotemark & ~\cite{NSR2005EL07} \\
$^{1}$H($^{17}$C, $^{17}$C+$\gamma$)    & $^{17}$C & 43.3 &             &      & & \\
$^{1}$H($^{17}$B, $^{17}$B+$\gamma$) & $^{17}$B & 43.8  & Liq.H$_{2}$ & 190  & d$\sigma$/d$\Omega$, $\delta$ & ~\cite{NSR2005DO16} \\
$^{1}$H($^{16}$C, $^{16}$C+$\gamma$) & $^{16}$C & 33.0  & Liq.H$_{2}$ & 225  & d$\sigma$/d$\Omega$, $\delta$ & ~\cite{NSR2006ON02} \\
$^{4}$He($^{22}$O, $^{23}$F+$\gamma$) & $^{22}$O & 35.0 & Liq.He & 100  & levels, d$\sigma$/d$\Omega$ & ~\cite{NSR2006MI16} \\
$^{4}$He($^{23}$F, $^{23}$F+$\gamma$) & $^{23}$F & 41.5 &        &      & & \\
$^{4}$He($^{24}$F, $^{23}$F+$\gamma$) & $^{24}$F & 36.0 &        &      & & \\
$^{4}$He($^{25}$Ne, $^{23}$F+$\gamma$) & $^{25}$Ne & 42.7 &      &      & & \\ 
CD$_{2}$($^{22}$O, $^{22}$O+$\gamma$) & $^{22}$O & 34.0  & CD$_{2}$ & 30  & $\sigma$, $\beta$ & ~\cite{NSR2006EL05} \\
$^{1}$H($^{30}$Na, $^{30}$Na+$\gamma$) & $^{30}$Na & 50.0  & Liq.H$_{2}$ & 210  & levels, $\sigma$, $\beta$ & ~\cite{NSR2006EL03} \\
$^{1}$H($^{31}$Na, $^{30,31}$Na+$\gamma$) & $^{31}$Na &    &             &      & & \\
$^{1}$H($^{33}$Mg, $^{33}$Mg+$\gamma$)    & $^{33}$Mg &    &             &      & & \\
$^{1}$H($^{34}$Mg, $^{33,34}$Mg+$\gamma$) & $^{34}$Mg &    &             &      & & \\
$^{1}$H($^{28}$Ne, $^{27,28}$Ne+$\gamma$) & $^{28}$Ne & 51.3 & Liq.H$_{2}$ & 210 & levels, $\sigma$ & ~\cite{NSR2006DO09} \\
Pb($^{26}$Ne, $^{26}$Ne+$\gamma$) & $^{26}$Ne & 54.0  & Pb & 230  & levels, d$\sigma$/d$\Omega$, $B$(E2) & ~\cite{NSR2007GI06} \\
$^{9}$Be($^{18}$C, $^{17}$C+$\gamma$) & $^{18}$C & 79.0 & Be & 370 & lifetime, $B$(M1) & ~\cite{NSR2008SU12} \\
$^{9}$Be($^{18}$C, $^{18}$C+$\gamma$) & $^{18}$C & 79.0 & Be & 370 & lifetime, $B$(E2) & ~\cite{NSR2008ON02} \\
$^{9}$Be($^{16}$C, $^{16}$C+$\gamma$) & $^{16}$C & 72.0 &    &     & & \\
Pb($^{76,80}$Ge, $^{76,80}$Ge+$\gamma$) & $^{76,80}$Ge & 37.0  & Pb & 175  & levels, $B$(E2;$2_{2}^{+} \rightarrow 0_{g.s.}^{+}$) & ~\cite{NSR2008IW03} \\
Pb($^{20}$C, $^{20}$C+$\gamma$) & $^{20}$C & 37.6 & Pb & 1445 & $\sigma$, $B$(E2) & ~\cite{NSR2009EL03} \\
$^{1}$H($^{20}$C, $^{20}$C+$\gamma$) &    & 41.4 & Liq.H$_{2}$ & 190 & & \\
$^{1}$H($^{60}$Cr, $^{60}$Cr+$\gamma$) & $^{60}$Cr & 42.0 & Liq.H$_{2}$ & 72  & $E(4_{1}^{+})$, $R(4^{+}/2^{+})$, $\sigma$, $\delta$ & ~\cite{NSR2009AO01} \\
$^{1}$H($^{62}$Cr, $^{62}$Cr+$\gamma$) & $^{62}$Cr & 39.0 &             &     & & \\
$^{1}$H($^{32}$Mg, $^{32}$Mg+$\gamma$) & $^{32}$Mg & 46.5  & Liq.H$_{2}$ & 157  & $E(4_{1}^{+})$, $R(4^{+}/2^{+})$, levels, d$\sigma$/d$\Omega$, $\delta$ & ~\cite{NSR2009TA08} \\
C($^{32}$Ne, $^{32}$Ne+$\gamma$) & $^{32}$Ne & 226.0 & C & 2540 & $E(2_{1}^{+})$ & ~\cite{NSR2009DO10} \\
C($^{33}$Na, $^{32}$Ne+$\gamma$) & $^{33}$Na & 245.0 &   &      & & \\
C($^{32}$Na, $^{31,32}$Na+$\gamma$) & $^{32}$Na & 230 & C & 2540 & levels & ~\cite{NSR2010DO05} \\
C($^{33}$Na, $^{33}$Na+$\gamma$) & $^{33}$Na &  -   &   &      & & \\
C($^{34}$Na, $^{33}$Na+$\gamma$) & $^{34}$Na & 250 &   &      & & \\
Pb($^{21}$N, $^{21}$N+$\gamma$) & $^{21}$N & 52.0 & Pb & 1445 & $\sigma$, $B$(E2) & ~\cite{NSR2010EL05} \\
$^{1}$H($^{21}$N, $^{19,21}$N+$\gamma$) &  & 48.1 & Liq.H$_{2}$ & 190 & & \\
Pb($^{28}$S, $^{28}$S+$\gamma$) & $^{28}$S & 53.0  & Pb & 348  & d$\sigma$/d$\Omega$, $B$(E2) & ~\cite{NSR2012TO06} \\
(CH$_{2}$)$_{n}$($^{32}$Mg, $^{32}$Mg+$\gamma$) & $^{32}$Mg & 190.0  & (CH$_{2}$)$_{n}$ & 2130 & d$\sigma$/d$\Omega$, $\beta$ & ~\cite{NSR2012LI45} \\
C($^{32}$Mg, $^{32}$Mg+$\gamma$)                &           &        & C & 2540 & & \\
C($^{44}$S, $^{40,42}$Si+$\gamma$) & $^{44}$S & 210.0 & C & 2540 & $E(4_{1}^{+})$, $R(4^{+}/2^{+})$, $\sigma$ & ~\cite{NSR2012TA20} \\
C($^{40}$S, $^{38}$Si+$\gamma$) & $^{40}$S &          &   &      & & \\
$^{1}$H($^{58}$Ti, $^{58}$Ti+$\gamma$) & $^{58}$Ti & 42.0  & Liq.H$_{2}$ & 72  & $E(2_{1}^{+})$, levels, $\sigma$, $\delta$ & ~\cite{NSR2013SU20} \\
$^{9}$Be($^{55}$Sc, $^{54}$Ca+$\gamma$) & $^{55}$Sc & 220  & Be & 1848  & $E(2_{1}^{+})$, levels& ~\cite{NSR2013ST20} \\
$^{9}$Be($^{56}$Ti, $^{54}$Ca+$\gamma$) & $^{56}$Ti & 230  &    &   & & \\
C($^{40}$Si, $^{38}$Mg+$\gamma$) & $^{40}$Si & 226 & C & 2540 & $E(4_{1}^{+})$, $R(4^{+}/2^{+})$& ~\cite{NSR2013DO22} \\
C($^{39}$Al, $^{38}$Mg+$\gamma$) & $^{39}$Al & 219 &   &      & & \\
C($^{37}$Al, $^{34,36}$Mg+$\gamma$) & $^{37}$Al & 247 &   &      & & \\
C($^{36}$Mg, $^{34}$Mg+$\gamma$) & $^{36}$Mg & 236 &   &      & & \\
\hline
\hline
\end{tabular}
}
\begin{flushleft}
\hspace{1em}
\small{
\begin{tabular}{rl}
\\
9& $\beta$: deformation parameter\\
\end{tabular}
}
\end{flushleft}
\end{table*}

\subsection{Detector configuration and mechanical structure}
\label{config-structure}

\begin{figure}
\begin{center}
\includegraphics[width=8.5cm]{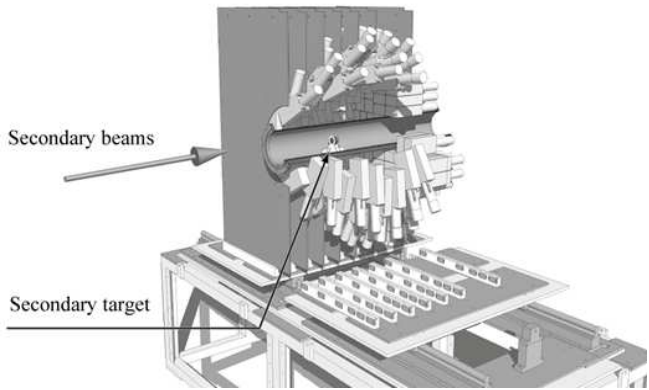}
\caption{
Schematic view of DALI2 in its standard configuration consisting of 186 NaI(Tl) crystals (see text for details).
}
\label{figure1}
\end{center}
\end{figure}

Figure~\ref{figure1} shows an illustration of DALI2 in its standard configuration 
that can accommodate 186 NaI(Tl) detectors.
Fast beams of unstable nuclei hit the target in a vacuum pipe, 
and DALI2 measures $\gamma$ rays from ejectiles produced in excited states, 
for example, by inelastic scattering, 
in coincidence with the ejectiles that are identified 
by a device such as the ZeroDegree Spectrometer~\cite{zds}.

DALI2 is composed of three types of NaI(Tl) scintillator crystals.
The first two types of crystals, 
which are manufactured by SAINT-GOBAIN~\cite{saintgobain} and SCIONIX~\cite{scionix}, 
have dimensions of 
$45\times80\times160$ mm$^3$ and $40\times80\times160$ mm$^3$, respectively. 
The third type of detector measures  
$60\times60\times120$ mm$^3$, and was fabricated originally 
for DALI by BICRON~\cite{saintgobain}. 
The crystals of the former two types are coupled to 38-mm$\phi$ 
HAMAMATSU~\cite{hamamatsu} R580 Photomultiplier tubes (PMTs), 
whereas the BICRON crystals use 50-mm$\phi$ HAMAMATSU R1306 PMTs.
Each crystal is encapsulated in a 1-mm-thick aluminum housing.  
The typical energy resolution is about 9\% (FWHM) 
for photons at 662 keV ($^{137}$Cs standard source).

\begin{figure}
\begin{center}
\includegraphics[width=8.5cm]{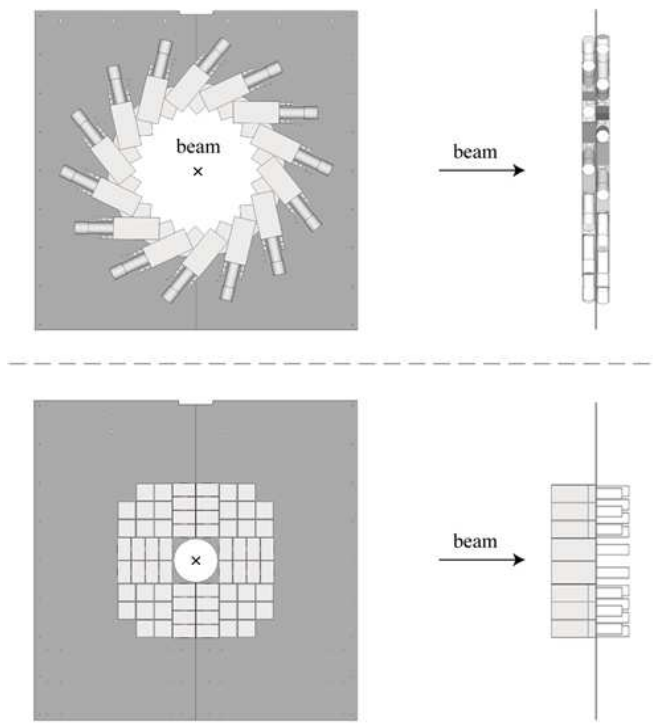}
\caption{
Top: layout of detectors in the 10$^{\rm th}$ and 11$^{\rm th}$ layers at 54 and 46 degrees, respectively. 
The other layers at backward angles are of similar configurations. 
Bottom: Detector configuration of the last layer for forward angle measurements.
}
\label{figure2}
\end{center}
\end{figure}

As shown in Fig.~\ref{figure1}, the detectors are arranged to form twelve layers set perpendicularly 
to the beam axis and a detector matrix covers the forward angles.
Each layer consists of 6-14 detectors,
which are mounted on a 5-mm-thick aluminum plate fixed to a support structure. 
For the 2$^{\rm nd}$ to 11$^{\rm th}$ layers, 
the two neighboring layers are attached to a single plate as shown in Fig.~\ref{figure2} (top).
The last plate, for the most forward angles, 
fixes the forward-angle matrix consisting of 64 crystals as shown in Fig.~\ref{figure2} (bottom).
The entire setup covers a range of polar angles in the laboratory frame 
between 15$^{\circ}$ and 160$^{\circ}$.
For the layers near 90$^{\circ}$, the distance between the target and the center of each detector is about 30 cm, 
which is shortest among the detectors in different layers.
In this standard DALI2 configuration, 
the opening angle spanned by a NaI(Tl) crystal is approximately 
6$^{\circ}$ (FWHM) for detectors at 60$^{\circ}$ to the beam line, 
where the Doppler-shift effect is largest for ejectiles with $v/c=0.6$.

For maintenance purposes, each layer is subdivided into two parts, 
and each part can be moved from its regular position.
The geometrical configuration of the crystals (or detectors) can be easily changed according 
to experimental requirements either by relocating detectors on 
the plates or by changing the layout of the plates.
This flexibility also enables different types of experiments, to be performed, 
such as life-time measurements using the recoil shadow method~\cite{NSR2008ON02}.

\subsection{Electronics}
\label{electronics}

\begin{figure}
\begin{center}
\includegraphics[width=8cm]{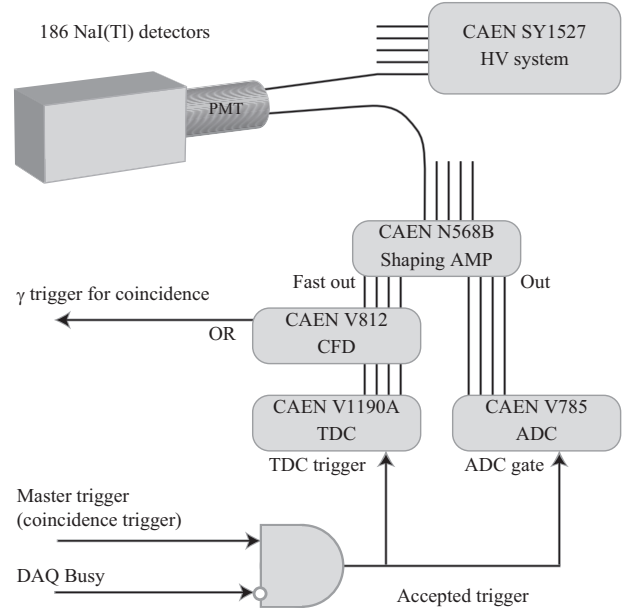}
\caption{
Schematic diagram of signal handling for DALI2.
}
\label{figure3}
\end{center}
\end{figure}

A typical diagram for signal handling for DALI2 is shown in Fig.~\ref{figure3}.
A multichannel power supply system (CAEN~\cite{caen} SY1527) 
provides a high voltage of around 1200 V to each PMT.
The signals from each PMT are fed into a shaping amplifier (CAEN N568B) 
and split into two separate signals for energy and timing measurements.
The signals for the energy measurements, outputs of N568B shaped with 3-$\mu$s time constant, 
are delivered to a peak-sensing analog-to-digital converter (CAEN V785 ADC).
Fast-out signals differentially shaped with 100-ns time constant are 
fed into a time-to-digital converter (CAEN V1190A TDC) 
through a constant-fraction discriminator (CAEN V812 CFD). 
To trigger the data acquisition, a logic signal ($\gamma$ trigger) 
is generated by taking the ``OR'' logic of the CFD outputs. 
This signal is also used as a gate signal for the ADC 
with 4 $\mu$s width and a trigger signal for the TDC. 
The V1190A TDC module has an adjustable time window 
with a time offset relative to the trigger signal, which are programmable via a VME control bus.
In most of the experiments, the width of time window and the offset are set to 2 $\mu$s and $-$1 $\mu$s,  
which enable to measure the $\gamma$-ray timing relative to the trigger signal, arrived about 1 $\mu$s later.
Thus, additional delays are not necessary for signals from CFDs. 
Data acquisition is performed by the system developed for experiments at RIBF~\cite{BABA10}.
For in-beam $\gamma$-ray spectroscopy experiments, 
the entire acquisition system 
is triggered by a coincidence signal generated by the ``AND'' logic of the $\gamma$-ray and beam triggers. 
A typical trigger rate is around 1 kHz 
and the dead time of the data acquisition is lower than 100 $\mu$s. 

\section{Performance}
\label{performance}

\subsection{Measurements of $\gamma$ transitions and energy resolution}
\label{require}

\begin{figure}[h]
\begin{center}
\includegraphics[width=7cm]{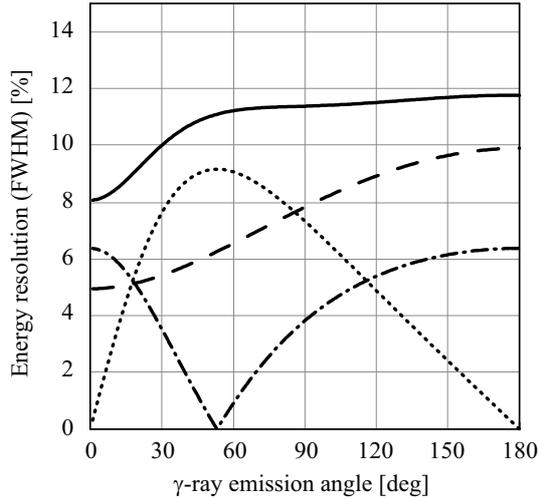}
\caption{
Expected energy resolution for 1-MeV photons in the center-of-mass system plotted 
as a function of $\gamma$-ray emission angle $\theta_{lab}$ (solid curve). 
The contributions from the intrinsic energy resolution (dashed curve), 
the finite opening angle of the detector (dotted curve), 
and the velocity spread of $\gamma$ emitters (dash-dotted curve) are also shown. 
See details in the text.
}
\label{figure4}
\end{center}
\end{figure}

As mentioned earlier, the resolution for the $\gamma$-transition energy is 
affected by the Doppler effect in addition to the intrinsic resolution of the detector. 
Experimental conditions such as target thicknesses, beam velocities, or beam directions also influence 
the energy resolution as described below.
The energy in the laboratory frame of reference, $E_{\rm lab}$, of $\gamma$ rays emitted from fast-moving nuclei 
is shifted from the $\gamma$-ray energy in the center-of-mass (CM) frame, $E_{\rm CM}$,  
i.e., the transition energy in the nucleus of interest. 
These energies are related as, 
\begin{eqnarray}
E_{\rm CM} =~ \gamma (1-\beta \cos\theta_{\rm lab}) E_{\rm lab},  
\label{eq1}
\end{eqnarray}
where 
$\theta_{\rm lab}$, $\beta$ and $\gamma$ denote respectively 
the $\gamma$-ray emission angle with respect to the direction of the moving nucleus,  
the velocity relative to the light speed ($v/c$) of the moving nucleus and the Lorentz factor.
Accordingly, the $E_{\rm CM}$ resolution depends on the uncertainties in 
the $\gamma$-ray angle, $\Delta \theta$, the beam velocity, $\Delta \beta$, 
and the measured $\gamma$ energy, $\Delta E_{\rm lab}$, according to 
\begin{eqnarray}
\left(\frac{\Delta E_{\rm CM}}{E_{\rm CM}}\right)^2 
&=~& 
\left(\frac{\beta \sin\theta_{\rm lab}}{1-\beta \cos\theta_{\rm lab}}\right)^2 
\left(\Delta \theta_{\rm lab}\right) ^2 \nonumber \\
&+~& 
\left(\frac{\beta\gamma^2(\beta-\cos\theta_{\rm lab})}
{1-\beta\cos\theta_{\rm lab}}\right)^2 
\left(\frac{\Delta \beta}{\beta}\right) ^2 \nonumber \\
&+~& 
\left(\frac {\Delta E_{\rm lab}}{E_{\rm lab}}\right) ^2. 
\label{eq2}
\end{eqnarray}
The angular resolution $\Delta \theta_{\rm lab}$ depends on the detector size and its distance from 
the $\gamma$-ray emission source. 
The uncertainty of the beam velocity $\Delta \beta$ is determined by the beam-energy spread and 
the thickness of the reaction target where the beam loses its energy. 
Figure~\ref{figure4} indicates the energy resolution for a 1-MeV $\gamma$ ray emitted from a nucleus with $\beta=0.6$ 
as a function of $\theta_{\rm lab}$.
Individual contributions from the three terms in Eq.~\ref{eq2} are shown separately. 
An intrinsic detector resolution $\Delta E_{\rm lab}/E_{\rm lab}$ of 7\% (FWHM), 
an averaged angular resolution ($\Delta \theta_{\rm lab}$) of 7 degrees (FWHM),  
and a velocity spread $\Delta\beta/\beta=10$\% are assumed, which are 
typical values for experiments at the new RIBF facility.
As seen in the figure, DALI2 is designed so that the overall contributions from the angular- 
and velocity-resolutions are similar to the NaI(Tl) intrinsic resolution, 
and hence a modest angular dependence of the $E_{\rm CM}$ resolution is achieved. 

\begin{figure}[h]
\begin{center}
\includegraphics[width=6.5cm]{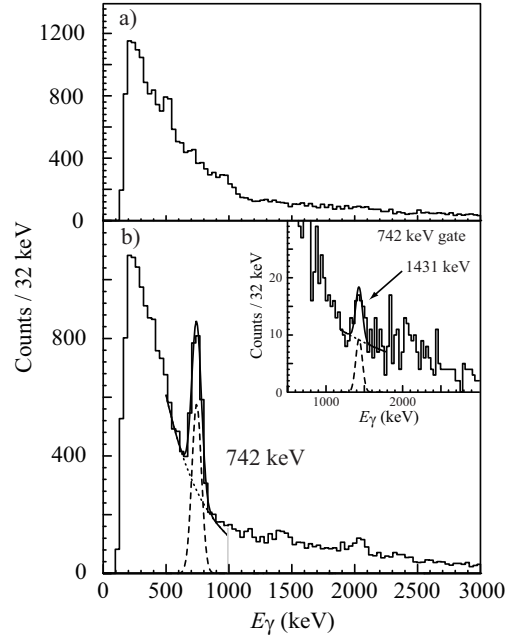}
\caption{
$\gamma$-ray energy spectra obtained for $^{42}$Si in the two-proton removal reaction experiments 
with fast RI beams of $^{44}$S~\cite{NSR2012TA20} without the correction for Doppler-shift effects (a) 
and with the correction (b). 
The inset of b) shows a $\gamma$-$\gamma$ coincidence spectrum gated on the 742-keV $\gamma$-ray line. 
The solid curves in b) represent the result of fits 
by using Gaussian function (dashed curves) and exponential background (dotted curves). 
}
\label{figure5}
\end{center}
\end{figure}

Figure~\ref{figure5} shows $\gamma$-ray energy spectra measured by DALI2 for the two-proton removal reaction 
feeding states in $^{42}$Si 
by radioactive $^{44}$S beams with $\beta=0.6$ and a carbon target (2.54 g/cm$^2$)~\cite{NSR2012TA20}. 
The spectrum in the laboratory frame, Fig.~\ref{figure5}(a), 
does not exhibit any distinct peaks expect for the one at 511 keV due to $e^+e^-$ annihilation.
The rise at low energies is caused by sources of atomic background~\cite{anhold84,anhold86,holdzmann93}. 
On the other hand, a few peaks are seen in the spectrum in the CM frame of $^{42}$Si Fig.~\ref{figure5}(b), 
obtained by correcting the Doppler shift using measured $\gamma$ emission angles.
The most prominent peak at 742 keV corresponds to the transition from the first excited 2$^{+}$state 
to the $0^+$ ground state in $^{42}$Si. 
Its width (91 keV at FWHM) is consistent with the above estimate and agrees well with Monte-Carlo simulations, 
which will be described in the following section. 
Other weaker peaks also correspond to transitions in $^{42}$Si. 
This is confirmed, for example, for the 1431-keV peak, which is more pronounced 
in the $\gamma$-$\gamma$ coincidence spectrum gated by the 742-keV $\gamma$ rays 
shown in the inset of Fig.~\ref{figure5}(b). 
As demonstrated in this example, 
DALI2 is capable of performing in-beam $\gamma$-ray spectroscopy experiments with secondary beams 
at very high velocities of $\beta \sim 0.6$.

\subsection{Detection efficiency}
\label{efficiency}

Table~\ref{tabeff} shows full-energy-peak efficiencies for $\beta$ values of 0.0, 0.3 and 0.6, 
at CM photon energies of 500~keV, 1~MeV and 2~MeV, 
obtained from Monte Carlo simulations using the code {\small GEANT3}~\citep{cern} 
with the standard configuration of DALI2 described in Sec.~\ref{config-structure} 
and an isotropic distribution of $\gamma$ rays. 
Effects of $\gamma$-ray absorption in material surrounding the target, the target holder 
(or the target cell in case of liquid target~\cite{crypta}) 
and the wall of the target chamber, for example, 
should be increased for more realistic measurements 
(they are not accounted for in the results listed in Table~\ref{tabeff}).
These effects reduce the efficiency typically by 20\% of the values listed in Table~\ref{tabeff}.

\begin{figure}[h]
\begin{center}
\includegraphics[width=7cm]{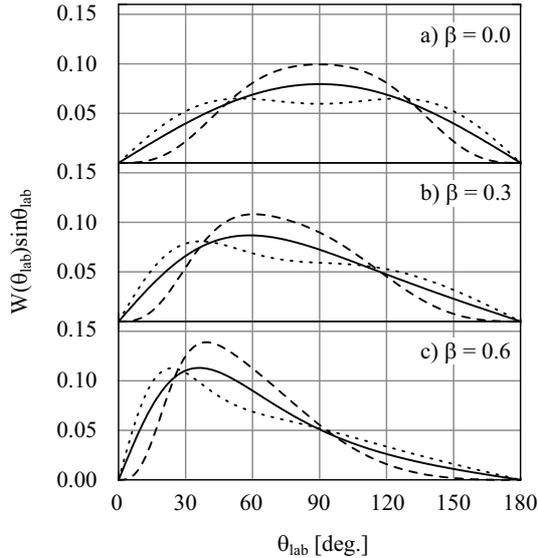}
\caption{
Angular distributions in the laboratory-frame for 1-MeV photons emitted from sources moving with velocities of 
a) $\beta=0.0$, b) $\beta=0.3$, and c) $\beta=0.6$.
The respective solid, dashed, and dotted curves show the distributions for isotropic $\gamma$ emission 
in the center-of-mass system, transitions with multipolarities of $\Delta L=1$ and $\Delta L=2$. 
Note that full spin-alignment is assumed for the cases of $\Delta L=1$ and $\Delta L=2$.
}
\label{figure6}
\end{center}
\end{figure}

Figure~\ref{figure6} shows the angular distribution for different $\beta$ and $\Delta L$ values, 
where the spin of the nucleus is assumed to be fully aligned 
in the plane perpendicular to the beam axis by the reaction (in the case of $\Delta L=1$ and $\Delta L=2$). 
DALI2 has an angular coverage of $\sim$15$^{\circ}$ to $\sim$160$^{\circ}$ with respect to the beam axis, 
resulting in a solid-angle coverage of more than 90\%.
Therefore, the uncertainty in the total $\gamma$ yield due to the incomplete angle-coverage is small. 
For example, the different assumptions for the multipolarities ($\Delta L=0,1,2$) 
and the velocities ($\beta =0.0, 0.3, 0.6$)
shown in Fig.~\ref{figure6} 
indicate a change of 6\% at most relative to the total yield.
The uncertainty might be even lower, 
since the alignment is lower in most cases.

\begin{figure}[h]
\begin{center}
\includegraphics[width=7cm]{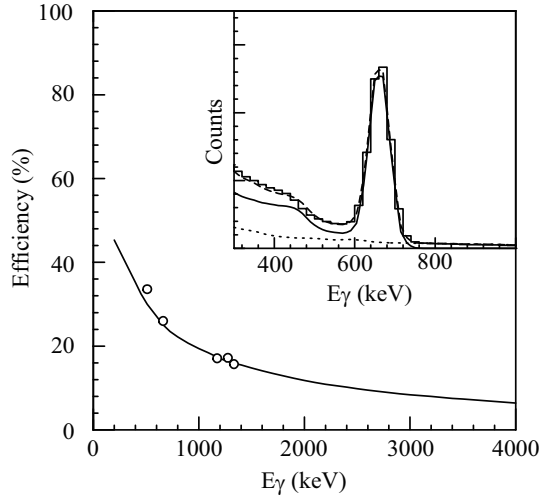}
\caption{
Full-energy-photopeak efficiencies measured with the standard sources $^{137}$Cs, $^{60}$Co and $^{22}$Na 
(open circles) and the one obtained by a simulation (solid curve).
The inset shows an energy spectrum for the $^{137}$Cs source using all the detectors of DALI2. 
}
\label{figure7}
\end{center}
\end{figure}

The Monte Carlo simulation was tested using measurements with standard stationary $\gamma$ sources,  
$^{22}$Na, $^{60}$Co and $^{137}$Cs, placed at the center of DALI2. 
In Fig.~\ref{figure7}, the results of a {\small GEANT3} simulation (solid curve) 
and measured efficiencies (open circles) are compared. 
They are consistent within 5\%, indicating a satisfactory accuracy of the simulation.
The inset of Fig.~\ref{figure7} shows a $\gamma$ spectrum taken with a $^{137}$Cs 
source using all the detectors of DALI2.
The dashed curve represents the simulated detector response to 662-keV photons. 
As shown in the figure, the experimental spectrum is reproduced well by the simulated response 
with an energy resolution of 9\% (mentioned in Sect.~\ref{config-structure}) and 
a component of measured background (dotted curve).

\begin{table}[b]
\caption{Simulated efficiencies of DALI2.}
\label{tabeff}
\center{
\begin{tabular}{c|ccc}
\hline
\hline
$E_\gamma$ (MeV) & $\beta=0.0$ & $\beta=0.3$ & $\beta=0.6$  \\
\hline
0.5 & 43\% & 40\% & 35\% \\
1.0 & 24\% & 24\% & 20\% \\
2.0 & 14\% & 13\% & 10\%\\
\hline
\hline
\end{tabular}
}
\end{table}

\subsection{Add-back analysis}
\label{addback}

A higher detection efficiency can be achieved by changing the detector configuration 
so that the average distance between the target and each detector 
is shorter than the standard setup described in Sect.~\ref{config-structure}.
However, this solution has a drawback:  
The $E_{\rm CM}$ resolution becomes worse because a larger opening angle is subtended by each detector, 
resulting in a more pronounced Doppler broadening effect.
An alternative way 
is to employ the so-called add-back analysis without changing the configuration.

In the add-back analysis, signals from a group of two or more neighboring detectors (a ``cluster'') 
were assumed to be generated from a Compton-scattered event. 
The sum of the energies measured in a given cluster is used to generate a $\gamma$-ray energy spectrum.
The detector with the largest energy deposit in the cluster 
is assumed to be the one where the first interaction takes place, 
and its position is used for the Doppler-shift correction.
By this add-back analysis, the photo-peak efficiency increases by 30\% 
for 1-MeV $\gamma$-rays emitted from nuclei moving at the velocity of $\beta=0.5$.
The possibility of misassignment of the first interaction point 
is found to be below 20\% of the cluster events using Monte Carlo simulations (under the same condition as above).
These events are taken into account as a part of the response function used for the fit.
In the proton inelastic scattering experiment to study $^{32}$Mg~\cite{NSR2009TA08}, 
the efficiency for 1-MeV $\gamma$ rays 
from moving nuclei with $v/c \sim 0.3$ was increased by about 20\% 
when the add-back analysis was adopted. 
In addition to increasing the detection efficiency, 
the add-back analysis improves the peak-to-background ratio, particularly for high-energy photons 
by reducing the Compton tails.
In order to obtain the location of a full-energy-peak in a $\gamma$ spectrum, 
the nonlinear response of the NaI(Tl) light output 
should be taken into account in the add-back analysis.  
The non-linearity is known to be sizable below 400 keV~\cite{knoll,leo}, 
and the resultant shift by this effect was estimated to be 5 keV in the analysis of 
the proton inelastic scattering on $^{32}$Mg~\cite{NSR2009TA08}. 

\subsection{$\gamma$-$\gamma$ coincidences}
\label{g-g}

The high efficiency of DALI2 enables $\gamma$-$\gamma$ coincidence measurements with reasonable statistics.
Reduction of the statistics compared with the case of the singles spectrum is by typically a factor of five 
for 1-MeV photons (see Table~\ref{tabeff}). 
It enables coincidence measurements to establish $\gamma$-ray cascades 
and hence level schemes of unstable nuclei produced with small yields. 
In the case of the two-proton removal reaction on $^{44}$S to $^{42}$Si~\cite{NSR2012TA20} 
described in Sect.~\ref{require}, 
the peak at 1431~keV becomes more pronounced in the coincidence spectrum gated on the 742-keV line 
relative to the singles spectrum, as shown in Fig.~\ref{figure5}(b) and its inset. 
This coincidence relationship for the 2$^+$ and 4$^+$ states in $^{42}$Si 
was supported by taking into account systematic trends for the states populated by the two nucleon removal reaction. 
This method has been applied successfully to establish low-lying level schemes, 
typically composed of 2$^+$ and 4$^+$ states, for many neutron-rich isotopes at RIBF, 
demonstrating the capability of the DALI2 array 
in in-beam $\gamma$-ray spectroscopy experiments for nuclei far from stability.

\subsection{Angular distributions}
\label{angular}

\begin{figure}[ht]
\begin{center}
\includegraphics[width=7.5cm]{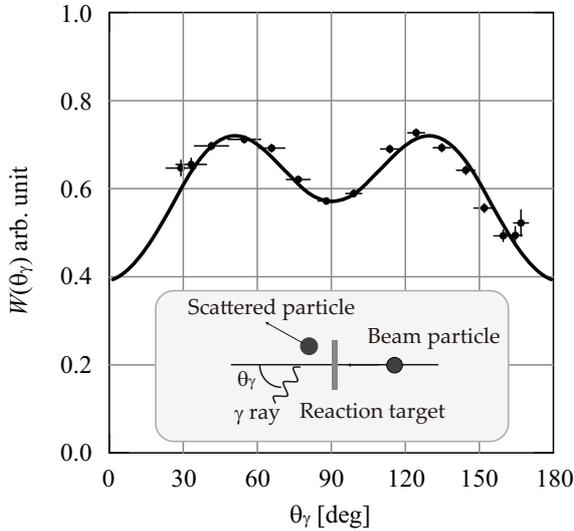}
\caption{
Angular distribution of 885~keV $\gamma$ rays emitted from moving $^{32}$Mg nuclei inelastically 
excited to the 2$^+$ state with a hydrogen target~\cite{NSR2009TA08}. 
The distribution is plotted in the rest frame of $^{32}$Mg.
The solid curve shows the result of a distorted-wave calculation for proton inelastic scattering assuming the multipolarity $\Delta L=2$. 
}
\label{figure8}
\end{center}
\end{figure}

As demonstrated in Fig.~\ref{figure6}, 
$\gamma$ rays from aligned nuclei have angular distributions that depend on transition multipolarities. 
Since a relatively large alignment is expected for the products of direct reactions at the RIBF energies, 
and therefore sizable sensitivity to the multipolarity is expected, 
the angular distribution measurements by DALI2 should provide information on the spins and parities 
of unknown excited states. 
Figure~\ref{figure8} shows an example. 
The angular distribution is for $\gamma$ rays from the first 2$^+$ excited state 
at $E_x = 885$ keV in $^{32}$Mg populated by inelastic proton scattering~\cite{NSR2009TA08}. 
The solid curve indicates the distribution, 
where the spin alignment was obtained from distorted-wave calculations 
assuming $\Delta L=2$ with the code {\small ECIS97}~\cite{ecis}.  
The theoretical angular distribution agrees well with the data taken by DALI2, 
indicating the usefulness of angular-distribution measurements in determining $\gamma$-ray multipolarities. 
The large angular coverage of DALI2 is ideal for this purpose. 

\begin{table}[b]
\caption{Basic parameters of DALI and DALI2. 
The efficiencies and energy resolutions are for 1-MeV $\gamma$ rays.}
\label{spec}
\center{
\begin{tabular}{c|cc|ccc}
\hline
\hline
&\multicolumn{2}{c|}{DALI}&\multicolumn{3}{c}{DALI2}\\
$\beta=v/c$& 0.0 & 0.3 & 0.0 & 0.3 & 0.6 \\
\hline
number of detectors & \multicolumn{2}{c|}{$\sim$60} & \multicolumn{2}{c}{160} & 186 \\
number of layers & \multicolumn{2}{c|}{6$\sim$8} & \multicolumn{3}{c}{16} \\
angle coverage (degree) & \multicolumn{2}{c|}{$\sim$50$^{\circ}$ -- $\sim$150$^{\circ}$} &  \multicolumn{3}{c}{$\sim$15$^{\circ}$ -- $\sim$160$^{\circ}$} \\
average $\Delta \theta$ (FWHM) & \multicolumn{2}{c|}{10$^{\circ}$} &   \multicolumn{3}{c}{7$^{\circ}$} \\
$\Delta E/E$ (FWHM) & 10\% & 12\% & 7\% & 8\% & 10\% \\
Efficiency & 13\% & 10\% & 24\% & 24\% & 20\% \\
\hline
\hline
\end{tabular}
}
\end{table}

\section{Discussion}

Table~\ref{spec} summarizes the efficiencies and resolutions of DALI and DALI2 for 1-MeV $\gamma$ rays 
emitted from $\gamma$-ray sources with different velocities. 
As mentioned in Sect.~\ref{intro}, 
DALI2 was designed to accommodate higher beam velocities at the new RIBF facility. 
As indicated in the table,  
the performance of 20\% efficiency and 10\% resolution at FWHM for DALI2 at $\beta=0.6$, 
the typical velocity of RI beams in the new RIBF facility, 
is preserved or even improved on compared to DALI at $\beta=0.3$, the velocity at RIPS in the old facility.
This resolution is satisfactory for spectroscopy 
of low-lying states in even-even exotic nuclei in light- and medium-mass regions, 
where their level spacings are typically larger than 500 keV. 
The high efficiency, in the $\sim$20\% range, enables $\gamma$-$\gamma$ coincidence measurements 
even for beam intensities as low as 1 Hz, making spectroscopic studies of nuclei very far from stability feasible. 

However, in order to extend our research opportunities to heavier exotic nuclei, 
such as the ones involved in the r-process nucleosynthesis, 
and to odd nuclei with narrower level spacings, 
a higher resolution device is needed.
An array called SHOGUN 
(Scintillator based High-resOlution Gamma-ray spectrometer for Unstable Nuclei)~\cite{shogun,shogun2} 
with LaBr$_3$(Ce) crystals was proposed for that purpose. 
The high intrinsic resolution of LaBr$_3$(Ce) ($\approx$ 2\% (FWHM) at 1.3~MeV is expected for a small-size crystal) 
is ideal for a higher energy-resolution array for fast-moving $\gamma$ emitters. 
The high $\gamma$-ray absorption coefficient of LaBr$_3$(Ce) material is 
another advantage for RI-beam induced experiments at RIBF. 
A design with one thousand crystals enables almost constant spread of Doppler shift for all the detectors 
by setting different distance from the target depending on the angle relative to the beam direction. 
The average opening angle of each detector is $\sim 3^{\circ}$. 
The expected energy resolution and efficiency for a 4$\pi$ geometry are 3.6\% (FWHM) 
and 40\%, respectively, for 1-MeV $\gamma$ rays emitted from nuclei moving at 200 MeV/u~\cite{shogun}. 
For more realistic estimate, 
the effect of velocity difference in the target discussed in Sect.~\ref{require} should be taken into account. 
For example, with a target causing beam-velocity change of 5\% (10\%), 
the overall energy resolution becomes 4.6\% (7.2\%).

Ge-based $\gamma$-ray trackers like GRETA~\cite{greta1,greta2} and AGATA~\cite{agata}  
can improve the energy resolution.
Their expected high angular resolution together 
with the high intrinsic resolution of Ge detectors 
can be exploited for applications to precision spectroscopy with a thin target. 
As already discussed, the target thickness, relevant to the velocity spread $\Delta \beta/\beta$, 
affects the energy resolution after correction of the Doppler-shift effect. 
In the condition with the target thickness causing 5\% change of the velocity of $\beta =0.5$, 
similar to the case discussed for SHOGUN, 
the expected energy resolution for 1-MeV $\gamma$ ray is to be 4.7\% for the AGATA array~\cite{agata-sim}, 
which is not different from the number for SHOGUN.
To achieve higher resolution, the velocity should be better controlled. 
For example, 
the velocity should be determined in 0.3\% precision with $\beta = 0.5$ 
to obtain energy-resolution values 10\% larger than the limiting value 
caused by other conditions as the intrinsic resolution~\cite{agata-sim}. 
The simplest solution is to use a very thin target, requiring high-intensity beams to keep the beam-target luminosity.

Ge-based $\gamma$-ray trackers are useful in spectroscopy experiments with fast beams of unstable nuclei 
not very far from the stability line, where the RI-beam intensity is expected to be high. 
On the other hand, 
in experiments for nuclei very far from stability line served as beams of a typical intensity of 1 Hz, 
use of scintillator-based arrays such as DALI2 or SHOGUN is a solution with good cost effectiveness.

\section{Summary}

We have developed a NaI(Tl) detector array, DALI2, 
for in-beam $\gamma$-ray spectroscopy experiments at RIKEN RIBF. 
The array consists of 186 NaI(Tl) detectors in its standard configuration. 
It subtends a large solid angle with a high efficiency. 
A typical full-energy-photopeak resolution of 10\% (FWHM) and 
20\% efficiency for photons with 1~MeV in the rest frame of 
the $\gamma$-ray emitter with $\beta \sim 0.6$ can be achieved.
The DALI2 array has been applied to 
various in-beam $\gamma$-ray spectroscopy experiments at RIBF successfully, 
and will be used in many more experiments 
to study structures of light and medium-mass exotic nuclei mostly with even numbers of protons and neutrons. 
The ZeroDegree~\cite{zds}, SAMURAI~\cite{samurai}, or SHARAQ~\cite{sharaq} spectrometers 
will be used to detect the reaction residues. 
For studies of heavy or odd-mass nuclei, 
the SHOGUN array, with superior energy resolution, is planned.

\section*{Acknowledgments}

The construction of DALI2 was supported financially by 
Research Center for Measurement in Advanced Science (RCMAS) of Rikkyo University 
and RIKEN Nishina Center for Accelerator-Based Science. 
We thank the RCMAS staff members and collaborators of experiments that were performed with DALI and DALI2. 
We are grateful to Dr. D.~Steppenbeck for proofreading of this article.



\begin{thebibliography}{76}
\expandafter\ifx\csname natexlab\endcsname\relax\def\natexlab#1{#1}\fi
\providecommand{\bibinfo}[2]{#2}
\ifx\xfnm\relax \def\xfnm[#1]{\unskip,\space#1}\fi
\bibitem[{{Metag} et~al.(1983)}]{metag}
\bibinfo{author}{V.~{Metag}}, et~al.,
\newblock \bibinfo{journal}{Nucl. Phys. A} \bibinfo{volume}{409}
  (\bibinfo{year}{1983}) \bibinfo{pages}{331c}.
\bibitem[{{Lee}(1990)}]{gammasphere}
\bibinfo{author}{I.-Y. {Lee}},
\newblock \bibinfo{journal}{Nucl. Phys. A} \bibinfo{volume}{520}
  (\bibinfo{year}{1990}) \bibinfo{pages}{641c}.
\bibitem[{{Beck}(1992)}]{euroball}
\bibinfo{author}{F.~A. {Beck}},
\newblock \bibinfo{journal}{Prog. Part. Nucl. Phys} \bibinfo{volume}{28}
  (\bibinfo{year}{1992}) \bibinfo{pages}{443}.
\bibitem[{{Simpson}(2000)}]{exogam}
\bibinfo{author}{J.~{Simpson}},
\newblock \bibinfo{journal}{APH N.S., Heavy Ion Physics} \bibinfo{volume}{11}
  (\bibinfo{year}{2000}) \bibinfo{pages}{159}.
\bibitem[{{Mueller} et~al.(2001)}]{sega}
\bibinfo{author}{W.~F. {Mueller}}, et~al.,
\newblock \bibinfo{journal}{Nucl. Instrum. Methods Phys. Res., Sect. A}
  \bibinfo{volume}{466} (\bibinfo{year}{2001}) \bibinfo{pages}{492}.
\bibitem[{{Lee} et~al.(2003)}]{greta1}
\bibinfo{author}{I.~Y. {Lee}}, et~al.,
\newblock \bibinfo{journal}{Rep. Prog. Phys.} \bibinfo{volume}{66}
  (\bibinfo{year}{2003}) \bibinfo{pages}{1095}.
\bibitem[{{Lee} et~al.(2004)}]{greta2}
\bibinfo{author}{I.~{Lee}}, et~al.,
\newblock \bibinfo{journal}{Nucl. Phys. A} \bibinfo{volume}{746}
  (\bibinfo{year}{2004}) \bibinfo{pages}{255c}.
\bibitem[{{Shimoura} et~al.(2004)}]{grape}
\bibinfo{author}{S.~{Shimoura}}, et~al.,
\newblock \bibinfo{journal}{Nucl. Instrum. Methods Phys. Res., Sect. A}
  \bibinfo{volume}{525} (\bibinfo{year}{2004}) \bibinfo{pages}{188}.
\bibitem[{{Akkoyun} et~al.(2012)}]{agata}
\bibinfo{author}{S.~{Akkoyun}}, et~al.,
\newblock \bibinfo{journal}{Nucl. Instrum. Methods Phys. Res., Sect. A}
  \bibinfo{volume}{668} (\bibinfo{year}{2012}) \bibinfo{pages}{26}.
\bibitem[{{Wisshak} et~al.(1990)}]{karlsruhe}
\bibinfo{author}{K.~{Wisshak}}, et~al.,
\newblock \bibinfo{journal}{Nucl. Instrum. Methods Phys. Res., Sect. A}
  \bibinfo{volume}{292} (\bibinfo{year}{1990}) \bibinfo{pages}{595}.
\bibitem[{{Weisshaar} et~al.(2010)}]{caesar}
\bibinfo{author}{D.~{Weisshaar}}, et~al.,
\newblock \bibinfo{journal}{Nucl. Instrum. Methods Phys. Res., Sect. A}
  \bibinfo{volume}{624} (\bibinfo{year}{2010}) \bibinfo{pages}{615}.
\bibitem[{{Warr} et~al.(2013)}]{miniball}
\bibinfo{author}{N.~{Warr}}, et~al.,
\newblock \bibinfo{journal}{Eur. Phys. J. A} \bibinfo{volume}{49}
  (\bibinfo{year}{2013}) \bibinfo{pages}{40}.
\bibitem[{RIB(2012)}]{RIBF}
\bibinfo{title}{{SPECIAL ISSUE: RESEARCH IN RI BEAM FACTORY}},
\newblock \bibinfo{journal}{Prog. Theor. Exp. Phys.}  (\bibinfo{year}{2012}).
\bibitem[{{Motobayashi} et~al.(1995)}]{NSR1995MO16}
\bibinfo{author}{T.~{Motobayashi}}, et~al.,
\newblock \bibinfo{journal}{Phys. Lett. B} \bibinfo{volume}{346}
  (\bibinfo{year}{1995}) \bibinfo{pages}{9}.
\bibitem[{{Nishio} et~al.(1996)}]{dali}
\bibinfo{author}{T.~{Nishio}}, et~al.,
\newblock \bibinfo{journal}{RIKEN Accel. Prog. Rep.} \bibinfo{volume}{29}
  (\bibinfo{year}{1996}) \bibinfo{pages}{184}.
\bibitem[{{Nakamura} et~al.(1997)}]{NSR1997NA08}
\bibinfo{author}{T.~{Nakamura}}, et~al.,
\newblock \bibinfo{journal}{Phys. Lett. B} \bibinfo{volume}{394}
  (\bibinfo{year}{1997}) \bibinfo{pages}{11}.
\bibitem[{{Iwasaki} et~al.(2000{\natexlab{a}})}]{NSR2000IW02}
\bibinfo{author}{H.~{Iwasaki}}, et~al.,
\newblock \bibinfo{journal}{Phys. Lett. B} \bibinfo{volume}{481}
  (\bibinfo{year}{2000}{\natexlab{a}}) \bibinfo{pages}{7}.
\bibitem[{{Iwasaki} et~al.(2000{\natexlab{b}})}]{NSR2000IW03}
\bibinfo{author}{H.~{Iwasaki}}, et~al.,
\newblock \bibinfo{journal}{Phys. Lett. B} \bibinfo{volume}{491}
  (\bibinfo{year}{2000}{\natexlab{b}}) \bibinfo{pages}{8}.
\bibitem[{{Yoneda} et~al.(2001)}]{NSR2001YO03}
\bibinfo{author}{K.~{Yoneda}}, et~al.,
\newblock \bibinfo{journal}{Phys. Lett. B} \bibinfo{volume}{499}
  (\bibinfo{year}{2001}) \bibinfo{pages}{233}.
\bibitem[{{Iwasaki} et~al.(2001)}]{NSR2001IW07}
\bibinfo{author}{H.~{Iwasaki}}, et~al.,
\newblock \bibinfo{journal}{Phys. Lett. B} \bibinfo{volume}{522}
  (\bibinfo{year}{2001}) \bibinfo{pages}{227}.
\bibitem[{{Iwasa} et~al.(2003)}]{NSR2003IW02}
\bibinfo{author}{N.~{Iwasa}}, et~al.,
\newblock \bibinfo{journal}{Phys. Rev. C} \bibinfo{volume}{67}
  (\bibinfo{year}{2003}) \bibinfo{pages}{064315}.
\bibitem[{{Shimoura} et~al.(2003)}]{NSR2003SH06}
\bibinfo{author}{S.~{Shimoura}}, et~al.,
\newblock \bibinfo{journal}{Phys. Lett. B} \bibinfo{volume}{560}
  (\bibinfo{year}{2003}) \bibinfo{pages}{31}.
\bibitem[{{Yanagisawa} et~al.(2003)}]{NSR2003YA05}
\bibinfo{author}{Y.~{Yanagisawa}}, et~al.,
\newblock \bibinfo{journal}{Phys. Lett. B} \bibinfo{volume}{566}
  (\bibinfo{year}{2003}) \bibinfo{pages}{84}.
\bibitem[{{Yamada} et~al.(2004)}]{NSR2004YA01}
\bibinfo{author}{K.~{Yamada}}, et~al.,
\newblock \bibinfo{journal}{Phys. Lett. B} \bibinfo{volume}{579}
  (\bibinfo{year}{2004}) \bibinfo{pages}{265}.
\bibitem[{{Elekes} et~al.(2004)}]{NSR2004EL03}
\bibinfo{author}{Z.~{Elekes}}, et~al.,
\newblock \bibinfo{journal}{Phys. Lett. B} \bibinfo{volume}{586}
  (\bibinfo{year}{2004}) \bibinfo{pages}{34}.
\bibitem[{{Imai} et~al.(2004)}]{NSR2004IM01}
\bibinfo{author}{N.~{Imai}}, et~al.,
\newblock \bibinfo{journal}{Phys. Rev. Lett.} \bibinfo{volume}{92}
  (\bibinfo{year}{2004}) \bibinfo{pages}{062501}.
\bibitem[{{Kondo} et~al.(2005)}]{NSR2005KO13}
\bibinfo{author}{Y.~{Kondo}}, et~al.,
\newblock \bibinfo{journal}{Phys. Rev. C} \bibinfo{volume}{71}
  (\bibinfo{year}{2005}) \bibinfo{pages}{044611}.
\bibitem[{{Iwasaki} et~al.(2005)}]{NSR2005IW02}
\bibinfo{author}{H.~{Iwasaki}}, et~al.,
\newblock \bibinfo{journal}{Phys. Lett. B} \bibinfo{volume}{620}
  (\bibinfo{year}{2005}) \bibinfo{pages}{118}.
\bibitem[{{Iwasa} et~al.(2008)}]{NSR2008IW04}
\bibinfo{author}{N.~{Iwasa}}, et~al.,
\newblock \bibinfo{journal}{Phys. Rev. C} \bibinfo{volume}{78}
  (\bibinfo{year}{2008}) \bibinfo{pages}{024306}.
\bibitem[{{Kondo} et~al.(2009)}]{NSR2009KO02}
\bibinfo{author}{Y.~{Kondo}}, et~al.,
\newblock \bibinfo{journal}{Phys. Rev. C} \bibinfo{volume}{79}
  (\bibinfo{year}{2009}) \bibinfo{pages}{014602}.
\bibitem[{{Elekes} et~al.(2004)}]{NSR2004EL10}
\bibinfo{author}{Z.~{Elekes}}, et~al.,
\newblock \bibinfo{journal}{Phys. Lett. B} \bibinfo{volume}{599}
  (\bibinfo{year}{2004}) \bibinfo{pages}{17}.
\bibitem[{{Kanungo} et~al.(2005)}]{NSR2005KA06}
\bibinfo{author}{R.~{Kanungo}}, et~al.,
\newblock \bibinfo{journal}{Phys. Lett. B} \bibinfo{volume}{608}
  (\bibinfo{year}{2005}) \bibinfo{pages}{206}.
\bibitem[{{Elekes} et~al.(2005)}]{NSR2005EL07}
\bibinfo{author}{Z.~{Elekes}}, et~al.,
\newblock \bibinfo{journal}{Phys. Lett. B} \bibinfo{volume}{614}
  (\bibinfo{year}{2005}) \bibinfo{pages}{174}.
\bibitem[{{Dombradi} et~al.(2005)}]{NSR2005DO16}
\bibinfo{author}{Z.~{Dombradi}}, et~al.,
\newblock \bibinfo{journal}{Phys. Lett. B} \bibinfo{volume}{621}
  (\bibinfo{year}{2005}) \bibinfo{pages}{81}.
\bibitem[{{Ong} et~al.(2006)}]{NSR2006ON02}
\bibinfo{author}{H.~J. {Ong}}, et~al.,
\newblock \bibinfo{journal}{Phys. Rev. C} \bibinfo{volume}{73}
  (\bibinfo{year}{2006}) \bibinfo{pages}{024610}.
\bibitem[{{Michimasa} et~al.(2006)}]{NSR2006MI16}
\bibinfo{author}{S.~{Michimasa}}, et~al.,
\newblock \bibinfo{journal}{Phys. Lett. B} \bibinfo{volume}{638}
  (\bibinfo{year}{2006}) \bibinfo{pages}{146}.
\bibitem[{{Elekes} et~al.(2006{\natexlab{a}})}]{NSR2006EL05}
\bibinfo{author}{Z.~{Elekes}}, et~al.,
\newblock \bibinfo{journal}{Phys. Rev. C} \bibinfo{volume}{74}
  (\bibinfo{year}{2006}{\natexlab{a}}) \bibinfo{pages}{017306}.
\bibitem[{{Elekes} et~al.(2006{\natexlab{b}})}]{NSR2006EL03}
\bibinfo{author}{Z.~{Elekes}}, et~al.,
\newblock \bibinfo{journal}{Phys. Rev. C} \bibinfo{volume}{73}
  (\bibinfo{year}{2006}{\natexlab{b}}) \bibinfo{pages}{044314}.
\bibitem[{{Dombradi} et~al.(2006)}]{NSR2006DO09}
\bibinfo{author}{Z.~{Dombradi}}, et~al.,
\newblock \bibinfo{journal}{Phys. Rev. Lett.} \bibinfo{volume}{96}
  (\bibinfo{year}{2006}) \bibinfo{pages}{182501}.
\bibitem[{{Gibelin} et~al.(2007)}]{NSR2007GI06}
\bibinfo{author}{J.~{Gibelin}}, et~al.,
\newblock \bibinfo{journal}{Phys. Rev. C} \bibinfo{volume}{75}
  (\bibinfo{year}{2007}) \bibinfo{pages}{057306}.
\bibitem[{{Suzuki} et~al.(2008)}]{NSR2008SU12}
\bibinfo{author}{D.~{Suzuki}}, et~al.,
\newblock \bibinfo{journal}{Phys. Lett. B} \bibinfo{volume}{666}
  (\bibinfo{year}{2008}) \bibinfo{pages}{222}.
\bibitem[{{Ong} et~al.(2008)}]{NSR2008ON02}
\bibinfo{author}{H.~J. {Ong}}, et~al.,
\newblock \bibinfo{journal}{Phys. Rev. C} \bibinfo{volume}{78}
  (\bibinfo{year}{2008}) \bibinfo{pages}{014308}.
\bibitem[{{Iwasaki} et~al.(2008)}]{NSR2008IW03}
\bibinfo{author}{H.~{Iwasaki}}, et~al.,
\newblock \bibinfo{journal}{Phys. Rev. C} \bibinfo{volume}{78}
  (\bibinfo{year}{2008}) \bibinfo{pages}{021304}.
\bibitem[{{Elekes} et~al.(2009)}]{NSR2009EL03}
\bibinfo{author}{Z.~{Elekes}}, et~al.,
\newblock \bibinfo{journal}{Phys. Rev. C} \bibinfo{volume}{79}
  (\bibinfo{year}{2009}) \bibinfo{pages}{011302}.
\bibitem[{{Aoi} et~al.(2009)}]{NSR2009AO01}
\bibinfo{author}{N.~{Aoi}}, et~al.,
\newblock \bibinfo{journal}{Phys. Rev. Lett.} \bibinfo{volume}{102}
  (\bibinfo{year}{2009}) \bibinfo{pages}{012502}.
\bibitem[{{Takeuchi} et~al.(2009)}]{NSR2009TA08}
\bibinfo{author}{S.~{Takeuchi}}, et~al.,
\newblock \bibinfo{journal}{Phys. Rev. C} \bibinfo{volume}{79}
  (\bibinfo{year}{2009}) \bibinfo{pages}{054319}.
\bibitem[{{Doornenbal} et~al.(2009)}]{NSR2009DO10}
\bibinfo{author}{P.~{Doornenbal}}, et~al.,
\newblock \bibinfo{journal}{Phys. Rev. Lett.} \bibinfo{volume}{103}
  (\bibinfo{year}{2009}) \bibinfo{pages}{032501}.
\bibitem[{{Doornenbal} et~al.(2010)}]{NSR2010DO05}
\bibinfo{author}{P.~{Doornenbal}}, et~al.,
\newblock \bibinfo{journal}{Phys. Rev. C} \bibinfo{volume}{81}
  (\bibinfo{year}{2010}) \bibinfo{pages}{041305}.
\bibitem[{{Elekes} et~al.(2010)}]{NSR2010EL05}
\bibinfo{author}{Z.~{Elekes}}, et~al.,
\newblock \bibinfo{journal}{Phys. Rev. C} \bibinfo{volume}{82}
  (\bibinfo{year}{2010}) \bibinfo{pages}{027305}.
\bibitem[{{Togano} et~al.(2012)}]{NSR2012TO06}
\bibinfo{author}{Y.~{Togano}}, et~al.,
\newblock \bibinfo{journal}{Phys. Rev. Lett.} \bibinfo{volume}{108}
  (\bibinfo{year}{2012}) \bibinfo{pages}{222501}.
\bibitem[{{Li} et~al.(2012)}]{NSR2012LI45}
\bibinfo{author}{K.~A. {Li}}, et~al.,
\newblock \bibinfo{journal}{Chin. Phys. Lett.} \bibinfo{volume}{29}
  (\bibinfo{year}{2012}) \bibinfo{pages}{102301}.
\bibitem[{{Takeuchi} et~al.(2012)}]{NSR2012TA20}
\bibinfo{author}{S.~{Takeuchi}}, et~al.,
\newblock \bibinfo{journal}{Phys. Rev. Lett.} \bibinfo{volume}{109}
  (\bibinfo{year}{2012}) \bibinfo{pages}{182501}.
\bibitem[{{Suzuki} et~al.(2013)}]{NSR2013SU20}
\bibinfo{author}{H.~{Suzuki}}, et~al.,
\newblock \bibinfo{journal}{Phys. Rev. C} \bibinfo{volume}{88}
  (\bibinfo{year}{2013}) \bibinfo{pages}{024326}.
\bibitem[{{Steppenbeck} et~al.(2013)}]{NSR2013ST20}
\bibinfo{author}{D.~{Steppenbeck}}, et~al.,
\newblock \bibinfo{journal}{Nature(London)} \bibinfo{volume}{502}
  (\bibinfo{year}{2013}) \bibinfo{pages}{207}.
\bibitem[{{Doornenbal} et~al.(2013)}]{NSR2013DO22}
\bibinfo{author}{P.~{Doornenbal}}, et~al.,
\newblock \bibinfo{journal}{Phys. Rev. Lett.} \bibinfo{volume}{111}
  (\bibinfo{year}{2013}) \bibinfo{pages}{212502}.
\bibitem[{{Takeuchi} et~al.(2003)}]{dali2}
\bibinfo{author}{S.~{Takeuchi}}, et~al.,
\newblock \bibinfo{journal}{RIKEN Accel. Prog. Rep.} \bibinfo{volume}{36}
  (\bibinfo{year}{2003}) \bibinfo{pages}{148}.
\bibitem[{dal(null)}]{dali2-web}
\bibinfo{title}{{The DALI2 website}},
  \bibinfo{howpublished}{\url{http://www.nishina.riken.jp/collaboration/SUNFLO%
WER/devices/dali2/index.html}}, \bibinfo{year}{\null}.
\bibitem[{{Ryuto} et~al.(2008)}]{crypta}
\bibinfo{author}{H.~{Ryuto}}, et~al.,
\newblock \bibinfo{journal}{Nucl. Instrum. Methods Phys. Res. Sect A}
  \bibinfo{volume}{590} (\bibinfo{year}{2008}) \bibinfo{pages}{204}.
\bibitem[{{Mizoi} et~al.(2005)}]{zds}
\bibinfo{author}{Y.~{Mizoi}}, et~al.,
\newblock \bibinfo{journal}{RIKEN Accel. Prog. Rep.} \bibinfo{volume}{38}
  (\bibinfo{year}{2005}) \bibinfo{pages}{297}.
\bibitem[{sai(null)}]{saintgobain}
\bibinfo{title}{{Saint-Gobain Crystals website}},
  \bibinfo{howpublished}{\url{http://www.detectors.saint-gobain.com}},
  \bibinfo{year}{\null}.
\bibitem[{sci(null)}]{scionix}
\bibinfo{title}{{SCIONIX Holland BV, P.O.Box 143, CC Bunnik, The
  Netherlands.}}, \bibinfo{year}{\null}.
\bibitem[{ham(null)}]{hamamatsu}
\bibinfo{title}{{Hamamatsu Photonics K.K.}},
  \bibinfo{howpublished}{\url{http://www.hamamatsu.com/jp/ja/index.html}},
  \bibinfo{year}{\null}.
\bibitem[{cae(null)}]{caen}
\bibinfo{title}{{CAEN S.p.A}},
  \bibinfo{howpublished}{\url{http://www.caen.it}}, \bibinfo{year}{\null}.
\bibitem[{{Baba} et~al.(2010)}]{BABA10}
\bibinfo{author}{H.~{Baba}}, et~al.,
\newblock \bibinfo{journal}{Nucl. Instrum. Methods Phys. Res. Sect. A}
  \bibinfo{volume}{616} (\bibinfo{year}{2010}) \bibinfo{pages}{65}.
\bibitem[{{Anhold} et~al.(1984)}]{anhold84}
\bibinfo{author}{R.~{Anhold}}, et~al.,
\newblock \bibinfo{journal}{Phys. Rev. Lett.} \bibinfo{volume}{53}
  (\bibinfo{year}{1984}) \bibinfo{pages}{234}.
\bibitem[{{Anhold} et~al.(1986)}]{anhold86}
\bibinfo{author}{R.~{Anhold}}, et~al.,
\newblock \bibinfo{journal}{Phys. Rev. A} \bibinfo{volume}{33}
  (\bibinfo{year}{1986}) \bibinfo{pages}{2270}.
\bibitem[{{Holzmann} et~al.(1993)}]{holdzmann93}
\bibinfo{author}{R.~{Holzmann}}, et~al.,
\newblock \bibinfo{journal}{GSI Annual Report 1992}  (\bibinfo{year}{1993})
  \bibinfo{pages}{48}.
\bibitem[{cer(1993)}]{cern}
\bibinfo{title}{{{\small GEANT3}, CERN Program Library Long Writeup W5013}},
  \bibinfo{howpublished}{\url{http://wwwasdoc.web.cern.ch/wwwasdoc/geant_html3%
/geantall.html}}, \bibinfo{year}{1993}.
\bibitem[{{Knoll}(1979)}]{knoll}
\bibinfo{author}{G.~F. {Knoll}}, \bibinfo{title}{{Radiation Detection and
  Measurement}}, \bibinfo{publisher}{Wiley, New York}, \bibinfo{year}{1979}.
  \bibinfo{note}{{ and references there in.}}
\bibitem[{{Leo}(1994)}]{leo}
\bibinfo{author}{W.~R. {Leo}}, \bibinfo{title}{{Techniques for Nuclear and
  Particle Physics Experiments}}, \bibinfo{publisher}{{Springer-Verlag,
  Berlin}}, \bibinfo{year}{1994}. \bibinfo{note}{{ and references there in.}}
\bibitem[{{Raynal}(null)}]{ecis}
\bibinfo{author}{J.~{Raynal}}, \bibinfo{title}{{coupled-channel code {\small
  ECIS97} (unpublished)}}, \bibinfo{year}{\null}.
\bibitem[{sho(null)}]{shogun}
\bibinfo{title}{{The SHOGUN website}},
  \bibinfo{howpublished}{\url{http://www.nishina.riken.jp/collaboration/SUNFLO%
WER/devices/shogun/index.html}}, \bibinfo{year}{\null}.
\bibitem[{{Doornenbal} et~al.(2009)}]{shogun2}
\bibinfo{author}{P.~{Doornenbal}}, et~al.,
\newblock \bibinfo{journal}{RIKEN Accel. Prog. Rep.} \bibinfo{volume}{42}
  (\bibinfo{year}{2009}) \bibinfo{pages}{182}.
\bibitem[{{Farnea} et~al.(2010)}]{agata-sim}
\bibinfo{author}{E.~{Farnea}}, et~al.,
\newblock \bibinfo{journal}{Nucl. Instrum. Methods Phys. Res., Sect. A}
  \bibinfo{volume}{621} (\bibinfo{year}{2010}) \bibinfo{pages}{331}.
\bibitem[{{Yoneda} et~al.(2012)}]{samurai}
\bibinfo{author}{K.~{Yoneda}}, et~al.,
\newblock \bibinfo{journal}{RIKEN Accel. Prog. Rep.} \bibinfo{volume}{45}
  (\bibinfo{year}{2012}) \bibinfo{pages}{i}.
\bibitem[{{Uesaka} et~al.(2008)}]{sharaq}
\bibinfo{author}{T.~{Uesaka}}, et~al.,
\newblock \bibinfo{journal}{Nucl. Instrum. Methods Phys. Res., Sect. B}
  \bibinfo{volume}{266} (\bibinfo{year}{2008}) \bibinfo{pages}{4218}.

\end{thebibliography}

\end{document}